\documentclass[]{emulateapj}

\usepackage{graphicx}
\usepackage[colorlinks=true,linkcolor=blue,citecolor=blue]{hyperref}

\shorttitle{Velocity shifts in protostellar jets}
\shortauthors{De Colle et al.}

\begin{document}


\title{Transverse velocity shifts in protostellar jets: rotation or velocity asymmetries?}


\author{Fabio De Colle \altaffilmark{1}, Adriano H. Cerqueira\altaffilmark{2}, Angels Riera\altaffilmark{3}}
\altaffiltext{1}{Instituto de Ciencias Nucleares, Universidad Nacional Aut\'onoma de M\'exico, A. P. 70-543 04510 D. F. Mexico \email{fabio@nucleares.unam.mx}}
\altaffiltext{2}{LATO-DCET-UESC, Rodovia Jorge Amado km 16, Ilh\'eus, Bahia, 45662-000, Brazil}
\altaffiltext{3}{Departament de F\'isica, EUETIB, Universitat Polit\'ecnica de Catalunya, Comte d'Urgell 187, E-08036, Barcelona, Spain }


\begin{abstract}
Observations of several protostellar jets show systematic differences 
in radial velocity transverse to the jet propagation direction, which have 
been interpreted as evidence of rotation in the jets. In this paper we discuss 
the origin of these velocity shifts, and show that they could be
originated  by rotation in the flow, or by side to side asymmetries
in the shock velocity, which could be due to asymmetries in the jet ejection
velocity/density or in the ambient medium. For typical poloidal jet velocities
($\sim 100-200$ km\,s$^{-1}$), an asymmetry $\gtrsim$ 10\% can produce
velocity shifts comparable to those observed. We also present three
dimensional numerical simulations of rotating, precessing and asymmetric jets,
and show that, even though for a given jet there is a clear degeneracy
between these effects, a statistical analysis of jets with different
inclination angles can help to distinguish between the alternative
origins of transverse velocity shifts. Our analysis indicate that
side to side velocities asymmetries could represent an important contribution
to transverse velocity shifts, being the most important contributor for large jet 
inclination angles (with respect the the plane of the sky), and can not be neglected when 
interpreting the observations.
\end{abstract}


\maketitle


\section{Introduction}
\label{sec1}

Astrophysical jets can be found in a variety of physical
scales, energies and environments. For instance, planetary nebulae,
cataclysmic variables, neutron stars and supernovae sometimes show
bipolar jets \citep[e.g.,][]{livio99}. In active galactic nuclei,
we can see accretion powered jets emanating from the central engine
\citep{yuan14}.  In star forming regions, jets are ubiquitous and
can be found associated with young stellar objects ranging from
brown dwarfs to high mass stars \citep[e.g.,][]{li14}.

Star forming regions like Orion and Taurus show a wealth of
protostellar jets.  For more than sixty years they have been studied
observationally, theoretically and, more recently, numerically.
These different approaches allowed to conclude
that protostellar jets are also accretion powered collimated
outflows, ejected episodically from the inner part of the accretion
disks with the help of a magnetic field. Supersonic variations in
the ejection velocity produce Herbig-Haro (HH) objects as emitting
post-shock cooling regions \citep[e.g.,][]{reipurth01}.  

Some of the HH objects have a knotty structure, while other have a 
bow-shaped form. It is still a matter of debate whether there 
is a contribution of a collimated stellar wind component, or a broad,
non-collimated disk wind component, and if the ejections are periodic
or not. Also, jets launched from the accretion disks are believed
to rotate \citep[e.g.,][]{ferreira06}. If measured, the rotation
of the jet with respect to its axis would give an estimation of the
amount of angular momentum extracted from the star-disk system. This is a key
parameter that can be used to understand the origin of protostellar
jets, and might also help to understand the general mechanism
responsible for the production of astrophysical jets at different
scales.

Recent observations of HH jets show transverse velocity shifts in
several emission lines.  \citet{davis00} first observed a shift in
the H$_2$ lines for the molecular jet HH\,212. \citet{bacciotti02}
observed the DG\,Tau micro-jet, much closer to the central source
and in atomic lines, finding a similar shift in several emission
lines. Other authors observed later a larger sample of objects
getting similar results: \citet{woitas05}, \citet{coffey04, coffey07,
coffey12} for class II objects, \citet{chrysostomou08} for Class I objects
 and \citet{lee07, lee08, choi11, coffey11} for class 0 
(or 0/I) objects.

From these results, important consequences on the jet ejection
mechanism have been inferred. Different models predict that jets
are ejected from different regions of the disk. For instance, in
the ``X-wind'' scenario the jet is ejected from the region of
interaction between the protostar's magnetosphere and the disk
\citep{shu00}, while in the ``disk-wind scenario'', the jet is
magnetocentrifugally ejected from an extended portion of the disk
surface \citep{blandford82}. 

If interpreted as rotation, observations
imply a range of ejection radii around $\sim 1$\,AU (e.g.
\citealt{ferreira06}), therefore excluding the X-wind as possible
mechanism for the jet ejection, and favoring the disk-wind scenario.
On the other side, recently \cite{lee08} determined wind launching
radii $\lesssim 0.05 - 0.30$\,AU, consistent with the X-wind model
\citep{shu00}\footnote{The dispersion in the disk wind
footpoint inferred from observations in different papers is mainly 
due to different poloidal terminal velocity used for the jet/outflow. 
The higher the poloidal velocity, the smaller the inner radius, as 
discussed in \cite{ferreira06}. \cite{bacciotti02} for instance have 
used $v_p = 80$ km\,s$^{-1}$ for the DG Tau jet, while \cite{lee08} 
have used $v_p \sim 100-200$ km\,s$^{-1}$ for HH\,212 molecular 
outflow, and they have found $r_0 \sim 1$ AU and $r_0 < 0.3$ AU, 
respectively.}.

In agreement with the rotation interpretation of the data, jet and
counter-jet rotate in the same sense in the Th\,28 and RW\,Aur jets.
In both cases, 
nevertheless, the sense of rotation observed in the
disk and the jet are opposite \citep{cabrit06, louvet16}.
In other cases the observations did not detect any rotation 
given the limits of resolution of the observations
(HH\,30 in \citealt{pety06} and \citealt{coffey07}, HH\,212 in \citealt{codella07},
RY\,Tau in \citealt{coffey15}). Puzzling, recent near-UV observations
\citep{coffey12} of the RW\,Aur jet showed velocity shifts which, 
if interpreted as rotation, give a sense of rotation consistent
with that of the disk (but opposite with respect
to previous optical observations), which were not detectable anymore
in following-up near-UV observations six months later 
at levels above the limiting resolution of the observations \citep{coffey12}.

While most of the studies have interpreted the presence of velocity
shifts as due to the rotation of the jet material, a few works have
studied alternative mechanisms. \citet[hereafter CER06]{cerqueira06},
using numerical simulations, showed that jet precession may lead
to velocity shifts similar to those observed, while \cite{soker05}
proposed disk asymmetries as a possible alternative mechanism. Among
the observed T Tau jets with velocity shifts, DG\,Tau has observed
precession \citep{dougados00, lavalley00} and the model developed
by CER06 may be applied to this jet.  \cite{laun09} have also
inferred, through radio observations, that the molecular outflow
associated with a T-Tauri star in the CB\,26 Bok globule (in
Taurus-Aurigae) is rotating. They have suggested that the outflow
is also precessing. For these cases, it seems plausible that the
effects of precession should be carefully taken into account when
considering radial velocity shifts as a fiducial evaluation for jet
rotation. 

More recently, \cite{pech2012} investigated the origin
of the radial velocity shift (of the order of 2 km s$^{-1}$) observed
in the HH\,797 outflow. In order to explain the observed data,
they have considered both precession and rotation, concluding that
rotation may probably account for the shifts. 
It seems at a first glance, however, that the HH\,797 jet is
precessing, as it is suggested by the wiggling of the outflow
far from the driving source \citep[see Figure 1 in][]{pech2012}.

In this paper we critically analyze the key hypothesis done implicitly
when interpreting the observed velocity shifts as rotation: the
absence of side to side (i.e., with respect to the jet axis) velocity
asymmetries.  In particular, we will show analytically (Section
\ref{sec2}) and by numerical models (Section \ref{sec3}) that the
presence of velocity asymmetries may generate effects resembling
those observed, i.e., that there is a degeneracy between toroidal
and asymmetric poloidal velocities, which can be disentangled only
by a statistical analysis of a large sample of HH jets.  In Section
\ref{sec4} we discuss the results, and in Section \ref{sec5} we
draw our conclusion.


\section{The origin of transverse velocity shifts in HH jets}
\label{sec2}

Observations of transverse velocity shifts (TVS) are commonly
interpreted as evidence of jet rotation. This is valid as long as
the dominant component in the radial velocity is the rotation
velocity. All the previous observational efforts done in order to
estimate the TVS are based on the assumption that
the jet velocity is symmetric across the jet radius, and that 
the TVS are only due to the presence of a given rotational profile.
Nevertheless, the poloidal jet velocity, which can have a large
variability even for a given jet \citep{ferreira06}, is a key point
to infer \citep[from the observational point of view; see, for
instance,][]{bacciotti02} the region at the surface of the accretion
disk where the outflow is actually produced/launched. 

A Herbig-Haro jet has a typical poloidal velocity of 100-200\,km\,s$^{-1}$
\citep{reipurth01}. The presence of a side to side (with respect to the jet axis) gradient
in the jet velocity (hereafter, a ``velocity asymmetry'') can also
contribute to the observed TVS.  To better illustrate this point,
in this section we will compute synthetic position-velocity diagrams
and, analyzing them similarly to how is done by observers, we will
demonstrate that rotation and velocity asymmetries produce similar
TVS. Then, we will show how it is possible to understand the origin
of TVS by a statistical analysis of the existing data.

\begin{figure}
\centering
\includegraphics[width=8cm]{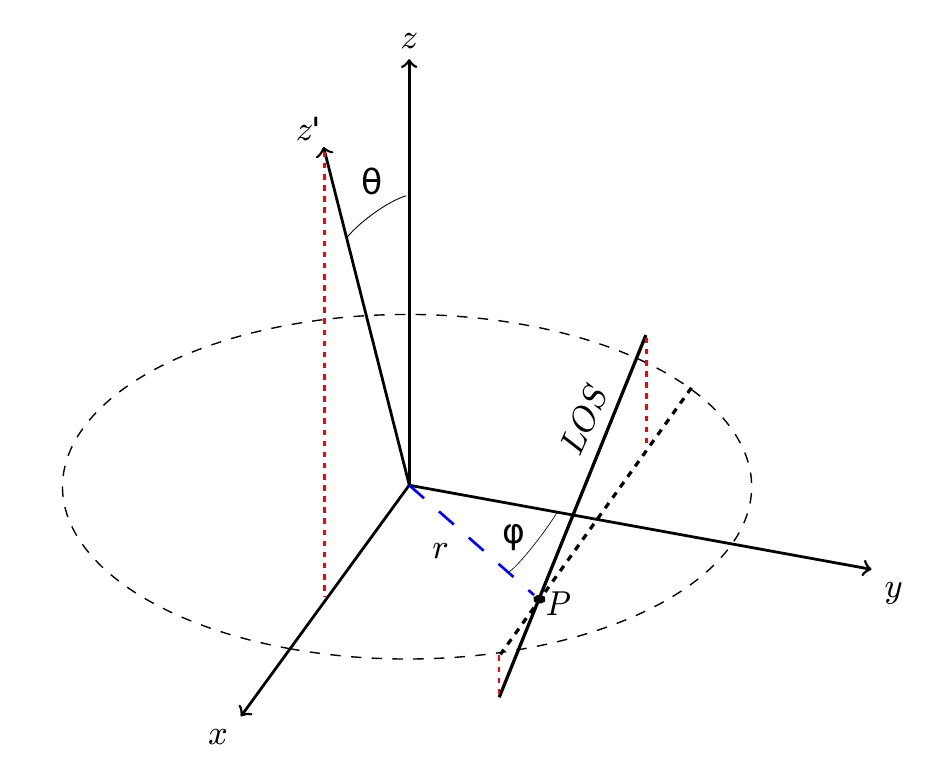}
\caption{Schematic representation of the jet geometry. The 
plane of the sky is represented by the $y,z^\prime$ plane. The 
jet moves along the $z$-axis, with an inclination angle $\theta$ 
with respect to the plane of the sky. The direction of the line of 
sight (LOS) is parallel to the planes $xz$, $xz^\prime$ and forms an 
angle $\theta$ with the $xy$ plane.}
\label{fig1}
\end{figure}

We consider a jet moving along the $z$-axis with velocity $v_z$ and
rotating around the $z$-axis with velocity $v_\phi$ (see figure
\ref{fig1}). Assuming for simplicity that the jet physical parameters
(i.e., jet density, temperature, velocity and chemical composition)
are independent of $z$, the radial velocity $v_r$ (along the line
of sight) of a fluid element $P$  is given by
\begin{equation}
  v_r(y,\phi) = v_\phi(r) \cos\phi \cos \theta + v_z(r,y,\phi) \sin \theta \;,
\label{eq:vr}
\end{equation}
where $\theta$ is the inclination angle of the jet with respect to
the plane of the sky, $\phi$ is the angle between the segment
connecting $P$ with the jet axis and the $y$-axis, $r$ is the
distance from $P$ to the jet axis, and $y=r\cos\phi$ (see figure
\ref{fig1}). 

The emission line intensity per unit velocity $I(y,v)$, observed
at a certain distance $y$ from the jet axis, can be computed by
integrating the emission coefficient per unit velocity $i(r,v)$
along the line of sight, i.e. by solving the ``Abel transform''\footnote{This 
equation neglects the convolution with seeing and instrumental
response (see the Appendix of \citealt{decolle10}).}:
\begin{equation}\label{eq:ixv}
 I(y,v) = \int_y^{\infty} \frac{i(r,v) r}{\sqrt{r^2-y^2}} dr \;.
\end{equation}

We assume that the emission coefficient per unit velocity $i(r,v)$
is related to the emission coefficient $i(r)$ by
\begin{equation}\label{eq:irv}
 i(r,v) = i(r)\;e^{-\frac{(v-v_r)^2}{\sigma^2}}\;,
\end{equation}
where $\sigma^2= \sigma^2_{\rm flow}+ \sigma^2_{\rm instr} $ is the
sum of the flow and the instrumental velocity dispersion. Typically
$\sigma^2_{\rm instr} \gg \sigma^2_{\rm flow}$, as e.g.  $\sigma^2_{\rm
instr}\approx 50$ km\,s$^{-1}$ for the ``Space Telescope Imaging
Spectrograph'' (STIS) on the Hubble Space Telescope (HST)\footnote{The
average dispersion per pixel for the G750M grating in the HST is
0.56 \AA ~ \citep{biretta16}. At H$\alpha$, this will give an
instrumental broadening of $\sim$ 25 km\,s$^{-1}$. Considering that
for an extended source this can be as twice as large, we will assume
$\sigma^2_{\rm instr}\approx 50$ km\,s$^{-1}$ \citep[see
also][]{hartigan09}.}.  Therefore, we can assume $\sigma^2 \approx
\sigma^2_{\rm instr}$.

By using equations \ref{eq:vr} and \ref{eq:irv}, equation \ref{eq:ixv} reduces to
\begin{equation}\label{eq:iyv}
 I(y,v) = \int_y^R \frac{i(r) e^{- (v - v_\phi y/r \cos \theta - v_z \sin \theta)^2 / \sigma^2}  r}{\sqrt{r^2-y^2}} dr \;.
\end{equation}

We assume a Gaussian dependence for the emissivity
$i(r)\propto\exp(-r/r_0)^2$ and a keplerian rotation velocity $v_\phi
= v_{\phi,0} (r/r_0)^{-1/2}$ (with $v_\phi = v_{\phi,0} (r/r_0)$
for $r<r_0$ to avoid an infinity at $r=0$).  The jet velocity
$v_z$ is in general a complicated function of $r,y,\phi$, e.g. a
decreasing function of $r$ and/or asymmetric with respect to $y$
and $\phi$. To focus only on the effects of asymmetries with respect
to the main axis of the jet (projected in the plane of the sky) we
take a simple jet velocity given by a modified ``top-hat'' profile,
i.e. $v_z = v_{z,0}(1-\Delta v_{z,0} x/R_j)$, being $R_j=1$ the
jet radius. Other velocity and intensity profiles will give
similar (at least qualitatively) results.

We use equation \ref{eq:iyv} to compute synthetic position-velocity
(PV) diagrams. From the PV diagrams we extract intensity profiles
at positions symmetric with respect to the jet axis.
A systematic gradient in the transverse Doppler profile is
what is usually interpreted as rotation in the observations.

\begin{figure}
  \centering
  \includegraphics[width=8cm]{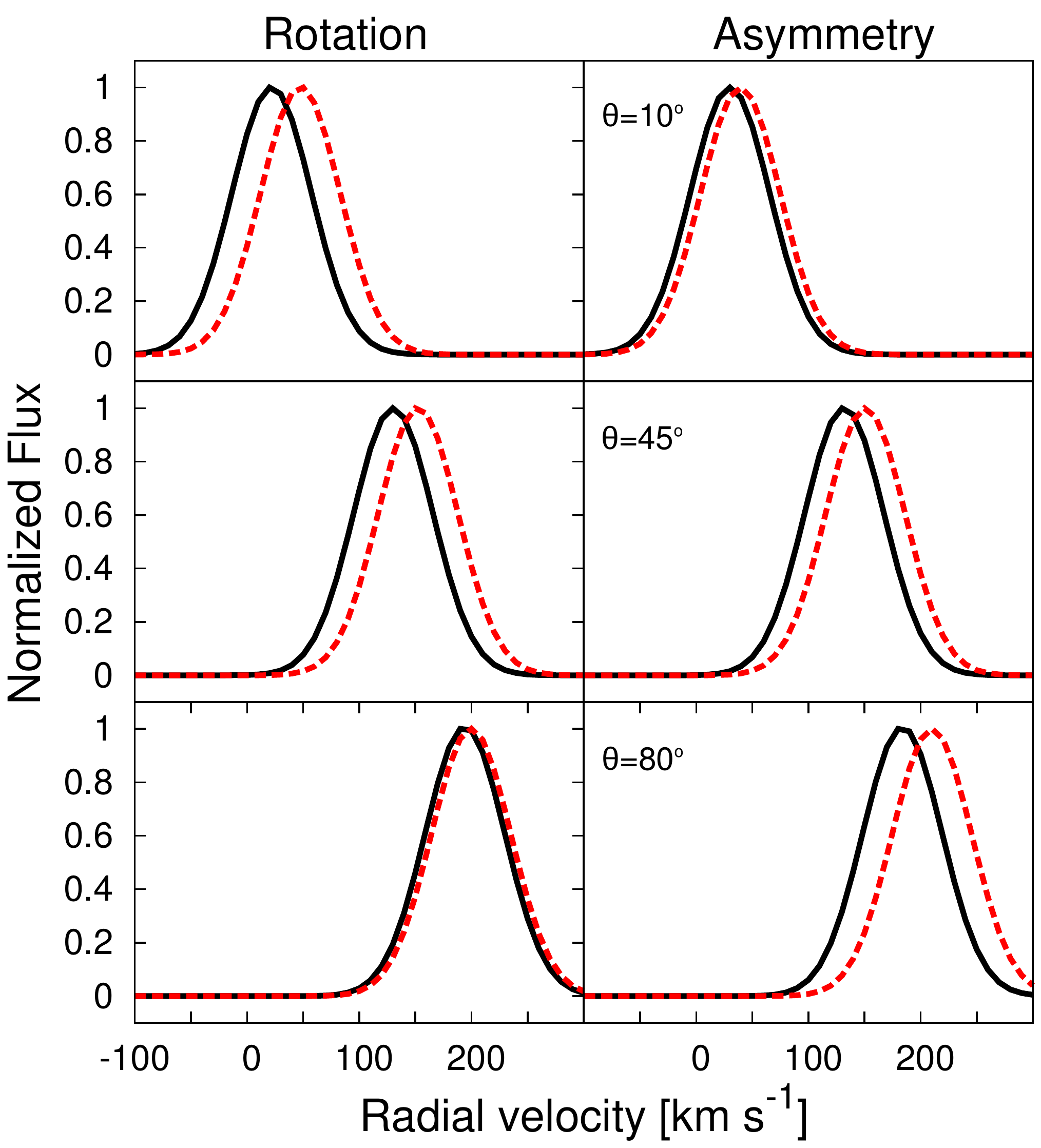}
\caption{Intensity profiles of synthetic emission lines computed in positions
symmetric with respect to the jet axis (black, full line and red,
dashed lines), for a jet with rotation  (left panels) and with
velocity asymmetries  (right), for inclination angles $\theta =
10^\circ, 45^\circ, 80^\circ$ (upper, middle, lower panels
respectively) with respect to the plane of the sky. The radial velocity differences, given by the shifts
in the synthetic emission lines, depend strongly on the jet inclination
angle. In particular, rotation (asymmetry) produces a larger velocity shift for small (large) jet inclination angles.}
   \label{fig2}
\end{figure}

Figure \ref{fig2} shows intensity profiles at $y=\pm \, 0.5\,R_j$,
for a rotating jet with $v_{\phi,0} = 200$ km\,s$^{-1}$, $r_0 =
0.001$ (corresponding to $v_{\phi} = 9$ km\,s$^{-1}$ at $r=0.5\,R_j$,
consistent with the toroidal velocity estimated by \citealt{coffey07}),
$\Delta v_{z,0} = 0$ (left panels), and for a jet with side to side
velocity asymmetries with $\Delta v_{z,0} = 0.1$, $v_{\phi,0} = 0$
(right panels). The poloidal jet velocity is equal to $v_{z_0} =
200$ km\,s$^{-1}$ in both cases.

Figure \ref{fig2} shows that unless the jet inclination angle is
$\theta \approx 0^\circ$ or $\theta \approx 90^\circ$, rotation and
velocity asymmetries both contribute to the radial velocity shift.
Therefore, observations of radial velocity shifts in a jet do
not allow to determine with precision the amount of rotation and
velocity asymmetry present in the jet, unless the jet is moving nearly 
in the plane of the sky.

Figure \ref{fig2} also shows that the TVS produced by rotation and velocity
asymmetry strongly depend on the jet inclination angle.  In fact,
the velocity gradient between two fluid elements located at positions
symmetric with respect to the jet axis (i.e. at $\phi=0$, $y =
y_0$ and $y = - y_0$) is given (see equation \ref{eq:vr}) by:
\begin{eqnarray} \label{eq:dvr}
 \Delta v_r &=& v_r (y_0,0) - v_r (-y_0,0) =   \nonumber \\
 &=& 2 v_\phi(r_0) \cos \theta + \Delta v_z \sin \theta
\end{eqnarray}
where $\Delta v_z = v_z (y_0,0) - v_z(-y_0,0)$ is the transverse gradient of the jet velocity.

If the jet is nearly axisymmetric, $\Delta v_z \ll v_\phi$. In this
case $\Delta v_r \sim 2 v_\phi \cos \theta$, and $\Delta v_r$
decreases for larger jet inclination angles.  If, on the other side,
the transverse velocity shift is mainly due to side to side asymmetries
in the poloidal component of the jet velocity ($\Delta v_z \gg
v_\phi$), $\Delta v_r \sim \Delta v_z \sin \theta$ and $\Delta v_r$
increases with the inclination angle of the jet.

In general we do not know \emph{a priori} which velocity component
dominates the radial velocity, but we can estimate the relative
importance of $v_\phi$ vs. $\Delta v_z$ by considering jets at
different inclination angles, as rotation (if present) will dominate
the observed velocity shifts at small inclination angles (i.e.,
$\Delta v_r \propto v_\phi$ if $\theta \ll \Delta v_z/v_\phi$, see
equation \ref{eq:dvr}), and velocity asymmetries (if present) will
dominate the velocity shifts at large inclination angles (i.e.,
$\Delta v_r \propto \Delta v_z$ if $\theta \gg v_\phi/\Delta v_z$).

Let us consider existing data of TVS determined observationally for
atomic lines.  The observations were presented by \citet{coffey04}
and \citet{coffey07} and include the CW\,Tau, DG\,Tau, HH\,30,
RW\,Aur and TH\,28 protostellar jets, observed by STIS on the Hubble
Space Telescope The HH\,30, CW\,Tau, DG\,Tau are located in the
Taurus molecular cloud at a distance of 140 pc, while Th\,28 and
RW\,Aur are located in Lupus 3 (140 pc) and Auriga (170 pc)
respectively. The estimated inclination angles are (see \citealt{coffey04,
coffey07} and references therein) 1$^\circ$ (HH\,30), 10$^\circ$
(Th\,28), 41$^\circ$ (CW\,Tau), 44$^\circ$ (RW\,Aur), 52$^\circ$
(DG\,Tau).

To estimate the relative importance of $v_\phi$ and $\Delta v_z$,
we computed, for each jet (see Table \ref{tab1}), the averages of
the transverse velocity shifts measured at different distances from
the jet axis by  \citet[][Table 3]{coffey04} and \citet[][Table
3]{coffey07} for the [\ion{O}{1}]$\lambda6300$, [\ion{N}{2}]$\lambda6583$,
[\ion{S}{2}]$\lambda 6716$, and [\ion{S}{2}]$\lambda6731$ emission
lines.  Other emission lines presented by \citet{coffey04, coffey07}
are not included as they are observed in a more limited number of
jets.

\begin{table}
\begin{center}
\scriptsize{\begin{tabular}{lcccc}
\hline \hline
 & $\Delta v_r $ for & $\Delta v_r$ for & $\Delta v_r$ for & $\Delta v_r$ for \\
Jet & [\ion{O}{1}]$\lambda$6300 & [\ion{N}{2}]$\lambda$6583 & [\ion{S}{2}]$\lambda$6716 & [\ion{S}{2}]$\lambda$6731 \\
 & (km~s$^{-1}$) & (km~s$^{-1}$) &(km~s$^{-1}$) & (km~s$^{-1}$) \\ 
\hline \hline
    HH\,30 blue   & 3.75 &  1.5 & -1.67& -2.4 \\
    TH\,28 red   & 13.4 & 16.5 &  3   &  1.2 \\
    TH\,28 blue & 11.5  & 6.67 &\dots & \dots \\ 
    CW\,Tau blue  &  14  & 13.5 &\dots &  -2  \\
    RW\,Aur red &15.25 &  14  & -1.5  & -1.33\\
    DG\,Tau blue & 17.6 & 6.67 & 2.33 & 2.6  \\
\hline \hline
\end{tabular}}
\end{center}
\caption{\scriptsize{Observed transverse velocity shifts}
\label{tab1}}
\end{table}

\begin{figure}
  \centering
  \includegraphics[width=8cm]{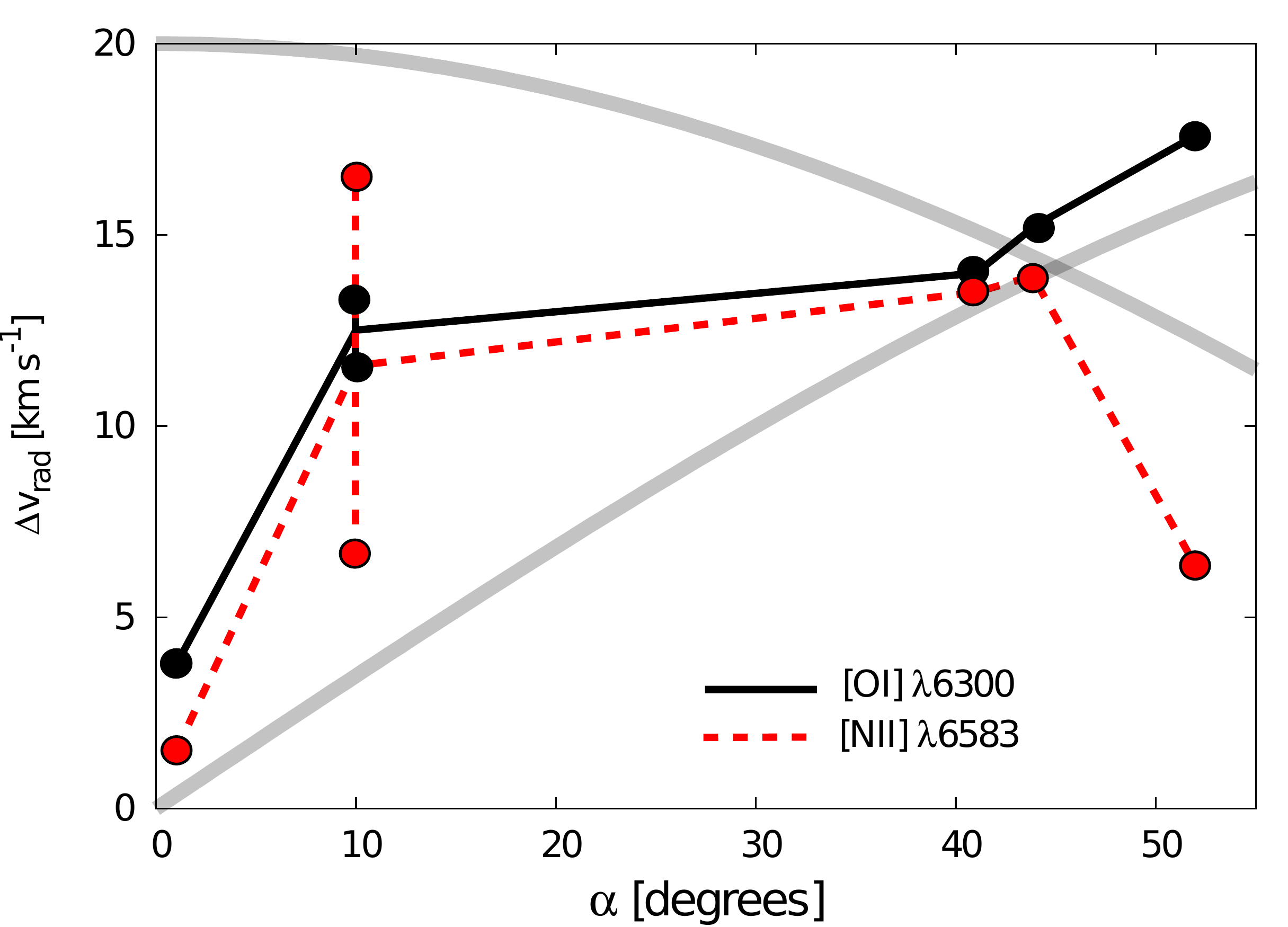}
  \caption{Comparison of observed shift in the transverse velocity with two simple model represented by $\sin \theta$ and $\cos \theta$ 
           curves for the velocity asymmetry and rotation cases. The data points 
           are from \citet{coffey04} and \citet{coffey07} and correspond 
           (from left to right) to HH\,30 (1$^\circ$), Th\,28 (10$^\circ$), CW\,Tau (41$^\circ$), RW\,Tau (44$^\circ$), DG\,Tau (52$^\circ$).}
   \label{fig3}
\end{figure}

In Figure \ref{fig3} we show the transverse velocity shift $\Delta
v_r$ as a function of the jet inclination angle for these protostellar
jets and the [\ion{O}{1}] and [\ion{N}{2}] emission lines.  The
other two lines considered do not show any measurable TVS (see Table
\ref{tab1}). For each emission line, we fit the data by using
equation \ref{eq:dvr}. To estimate the accuracy of the fit we 
have computed the parameter $Q$, which
is the probability that, given the fit to the data, data with
Gaussian noise (assumed here to be 5 km\,s$^{-1}$) have a $\chi^2$
larger than the one determined in the fit. Values of $Q\lesssim 1$
indicates an acceptable fit, while smaller values (Q $\approx 0$)
indicate that the fit is poor.

The results are the following: for a purely rotating jet, $\Delta
v_r = 2 v_\phi \cos \theta$, with $v_\phi = 6.9 \pm 1.6$ km\,s$^{-1}$,
$Q=0.1$ for the [\ion{O}{1}] line, and $v_\phi = 5.5 \pm 1.5$
km\,s$^{-1}$, $Q=0.15$ for the [\ion{N}{2}] line.  For a jet with
velocity asymmetry we get $\Delta v_r = \Delta v_z \sin \theta$
with $\Delta v_z = 23.8 \pm 4.4$ km\,s$^{-1}$, $Q=0.28$ for the
[\ion{O}{1}] line, and $\Delta v_z = 17.5 \pm 5.6$ km\,s$^{-1}$,
$Q=0.07$ for the [\ion{N}{2}] line. The best fit is obtained by
including both rotation and velocity asymmetries, with  $v_\phi=3.7\pm
1.0$ km\,s$^{-1}$, $\Delta v_z = 15.4\pm 3.2$ km\,s$^{-1}$, $Q=0.8$
for the [\ion{O}{1}] line, and $v_\phi=3.6\pm 1.9$ km\,s$^{-1}$,
$\Delta v_z = 9.1\pm 6.4$ km\,s$^{-1}$, $Q=0.25$ for the [\ion{N}{2}]
line.

Fits to the data give typical velocities $v_\phi \sim 5$ km\,s$^{-1}$
and $\Delta v_z \sim 15$ km\,s$^{-1}$. Although this result should
be taken carefully because of the low statistics and the approximations
used (e.g., we are assuming that $v_\phi$ and $\Delta v_z$ are the
same for all jets, and this is not necessarily true), it seems to
indicate that velocity asymmetries are an important component of
the velocity shifts, dominant for jets with ``large'' inclination
angles ($\theta \gtrsim \arctan (2v_\phi/\Delta v_z)$, see equation
\ref{eq:dvr}).

\section{Numerical simulations}
\label{sec3}

When the jet inclination is considered, different shock regions,
located at different positions $z$ along the direction of propagation
of the jet, contribute to the radial velocity.  In this Section,
we will use three dimensional numerical simulations of protostellar
jets to take in account multi-dimensional effects neglected in
Section \ref{sec2}.

We have used the Yguaz\'u-a code to simulate jets in three-dimensions.
The Yguaz\'u-a is an adaptive grid code that uses a flux vector
splitting scheme \citep[see][]{vanalbada82} to evolve the equations
of hydrodynamics. At each time step, the code integrates a system
of rate equations for 17 atomic/ionic species, namely: \ion{H}{1},
\ion{H}{2}, \ion{He}{1}, \ion{He}{2}, \ion{He}{3}, \ion{C}{2},
\ion{C}{3}, \ion{C}{4}, \ion{N}{1}, \ion{N}{2}, \ion{N}{3}, \ion{O}{1},
\ion{O}{2}, \ion{O}{3}, \ion{O}{4}, \ion{S}{2}, \ion{S}{3}. A
non-equilibrium cooling function is then calculated. The detailed
reaction equations as well as the cooling function are given in
\citet{raga00, raga07}.

The number of cells in the computational box is ($x$, $y$, $z$) =
(128, 128, 512). Each cell has a physical dimension of 8.59$\times10^{13}$
cm, or 5.7 AU, for all the three dimensions (in a Cartesian coordinate
system).  A jet with a radius of $R_j = 37.4$ AU\footnote{This
value for the jet radius is suggested by the observations of DG Tau
micro-jet presented by Bacciotti et al (2001), who placed seven slits
across the jet axis, separated by a distance of $\sim 10$ AU to
cover all the emitting region. This gives a jet diameter estimative
of $\sim$ 60 AU, and then $R_j \sim 30$ AU.} is injected from the
$xy$ plane at $z = 0$ (see Figure \ref{fig1}), and propagates in
the positive $z$ direction. The jet to ambient medium density ratio
is given initially by $n = n_j/n_a = 10$, where $n_j = 1\,000$
cm$^{-3}$. The initial jet ionization fraction of H is 0.1. The jet
temperature is $T_j = 10^4$ K, and the ambient medium temperature
is $T_a = 10^3$ K. The initial setup is equivalent to the one
employed in CER06.

We present here the results from four different numerical simulations,
differing for the presence (or not) of rotation, precession and
side to side velocity asymmetry (see Table \ref{tab2}). The baseline
jet velocity (in the $z$ direction) is the same for all the models:
300 km\,s$^{-1}$. Model M1 does not have rotation while the others
do have. The model M4 has also a precession (see below).  All models
have a sinusoidally variable jet velocity (along the $z$ direction):

\begin{equation} 
V_j \equiv V_z =  V_0 \cdot \bigg[ 1 + A\,{\rm sin} \bigg( \frac{2 \pi}{P} t \bigg) 
\bigg] \cdot [1 + f(y)]
\end{equation}

\noindent where $V_0 = 300$ km\,s$^{-1}$, $A = 0.33$ and $P = 8$
yr (these parameters are chosen to reproduce the velocity
structure of the DG Tau micro-jet; \citealt{dougados00,raga01}).
Models M2, M3 and M4 have $f(y) = 0$. This is the typical top-hat
profile for the $V_z$ adopted in several numerical studies of jets.
The model M1 has $f(y) = 0.025\; {\rm sign}(y)$, where ${\rm sign}(y)$
is a function  that returns the sign of the $y$ coordinate in the
computational domain. This implies that at any time step $\Delta
V_z(t) = 0.05\;V_j(t)$ in the $yz$ plane. We could expect, then, a
maximum side to side jet asymmetry of $\Delta V_z = 20$ km\,s$^{-1}$.
We are artificially  introducing an asymmetry between both sides
of the jet axis in the jet velocity, without arguing about its
nature. Although quite speculative, this model will serve to
illustrate what can actually occur if a jet has, for some reason,
a side to side asymmetry in shock velocities.

\begin{table}
\begin{center}
\scriptsize{\begin{tabular}{lccc}
\hline\hline
Model     & $\tau_{\rm prec}$ & $V_{\phi} \times (r/R_j)$ & $\Delta V_z$ \\
          & (years)       & (km\,s$^{-1}$)         & (km\,s$^{-1}$)   \\
\hline
M1        &    0          & 0          &  $\leq$ 20   \\
M2        &    0          & $\lesssim$ 10       &  0               \\
M3        &    0          & 8         &  0               \\
M4        &    8          & 8         &  0               \\
\hline
\hline
\end{tabular}}
\end{center}
\caption{\scriptsize{The simulated models}
\label{tab2}}
\end{table}

M2, M3 and M4 are rotating jet models with different rotational
velocity profiles ($V_{\phi} = \sqrt{V_x^2 + V_y^2}$). In model M2
$v_{\phi}$ is given by:

\begin{equation}
V_{\phi} = \frac{V_j}{40}
\end{equation}

\noindent so the rotational velocity will fluctuate in time with
values ranging from 5 and 10 km\,s$^{-1}$ (it is constant through
the jet cross section). The model M3 is the same model that we have
presented in CER06. For this case:

\begin{equation}
V_{\phi} = 8~{\rm km~s}^{-1}\cdot \frac{R_j}{r}
\end{equation}

\noindent where $r = \sqrt{x^2 + y^2}$.  As in CER06, profile has
been truncated at $R = 0.15 R_j$. The toroidal velocity ranges then
from $V_{\phi} = 55$ km\,s$^{-1}$ to 8 km\,s$^{-1}$ at the jet
radius \citep[see also][]{cerqueira04}. The model M4 is similar to
M3, but the jet is actually precessing with a half opening angle
of 5$^{\circ}$ and with a precessional period of $\tau_{\rm prec}
= 8$ years.

The strategy that we have used here in order to build the synthetic
slits, the synthetic line profiles and to analyze the line profile
in terms of its components (low-, medium- and high-velocity component)
is the same that we have presented in CER06.  Firstly, we calculate
the emission coefficients for the [\ion{O}{1}] and [\ion{S}{2}]
emission lines \citep{raga04}. To build velocity channel maps (VCM) for a given
radial velocity $V_r$, the local emissivity is smeared out (in
radial velocity, or wavelength) using a Gaussian profile. The
broadening of the line profile is estimated using the characteristic
sampling (we use $\Delta V_r = 10$ km\,s$^{-1}$) of our VCMs
and the local sound speed, $c_s$. We then define the local dispersion
of the line profile as $\sigma = \sqrt{\Delta V_r^2 + c_s^2}$.

To build the line profiles we need to define an artificial slit,
which depends on  the position in the VCMs, and we need to
project our computational box on the plane of the sky.  
The radial velocity is then the projection of the
velocity at a given cell along the line of sight, which can make
an arbitrary angle with the $x-$axis. Increasing the inclinations
angles will shift the profiles to increasingly negative radial
velocities.

\begin{figure}
\centering
\includegraphics[width=8cm]{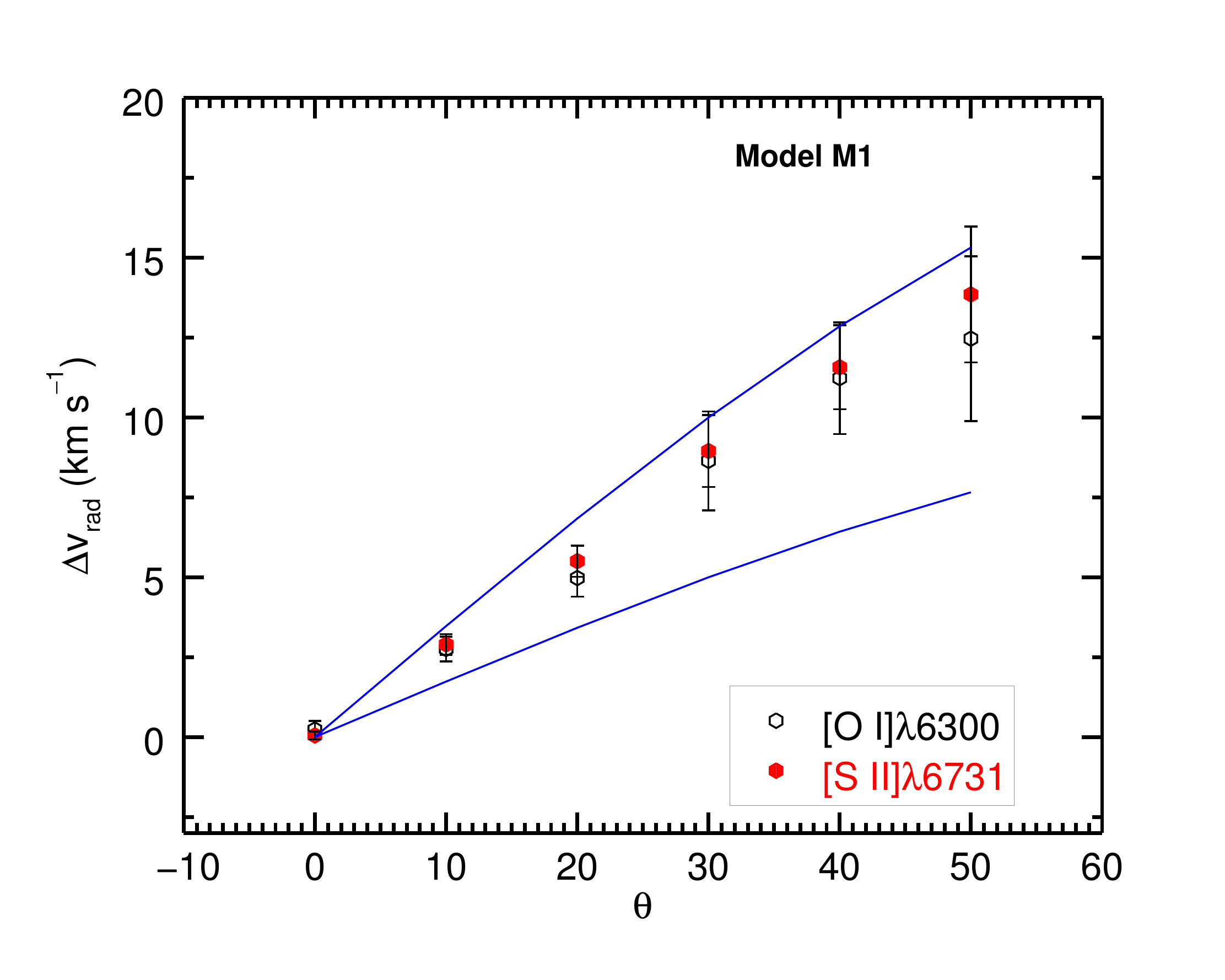}
\caption{Radial velocity estimates from different line profiles
(filled circles, [\ion{S}{2}]6731; open circles, [\ion{O}{1}]6300),
and for different inclination angles from 0$^{\circ}$ to $50^{\circ}$,
for the model M1 (i.e., the model with side to side shock velocity
asymmetries). For each inclination angle, we have used four different
regions along the jet, and six different positions across the jet
axis to compute $\Delta v_{\rm rad}$. The dispersion in the radial velocity around its mean value
is indicated by the error bars for each inclination angle. The solid
curves are sine functions with amplitude of 10 and 20
km\,s$^{-1}$.}
\label{fig4}
\end{figure}

\begin{figure}
\centering
\includegraphics[width=8cm]{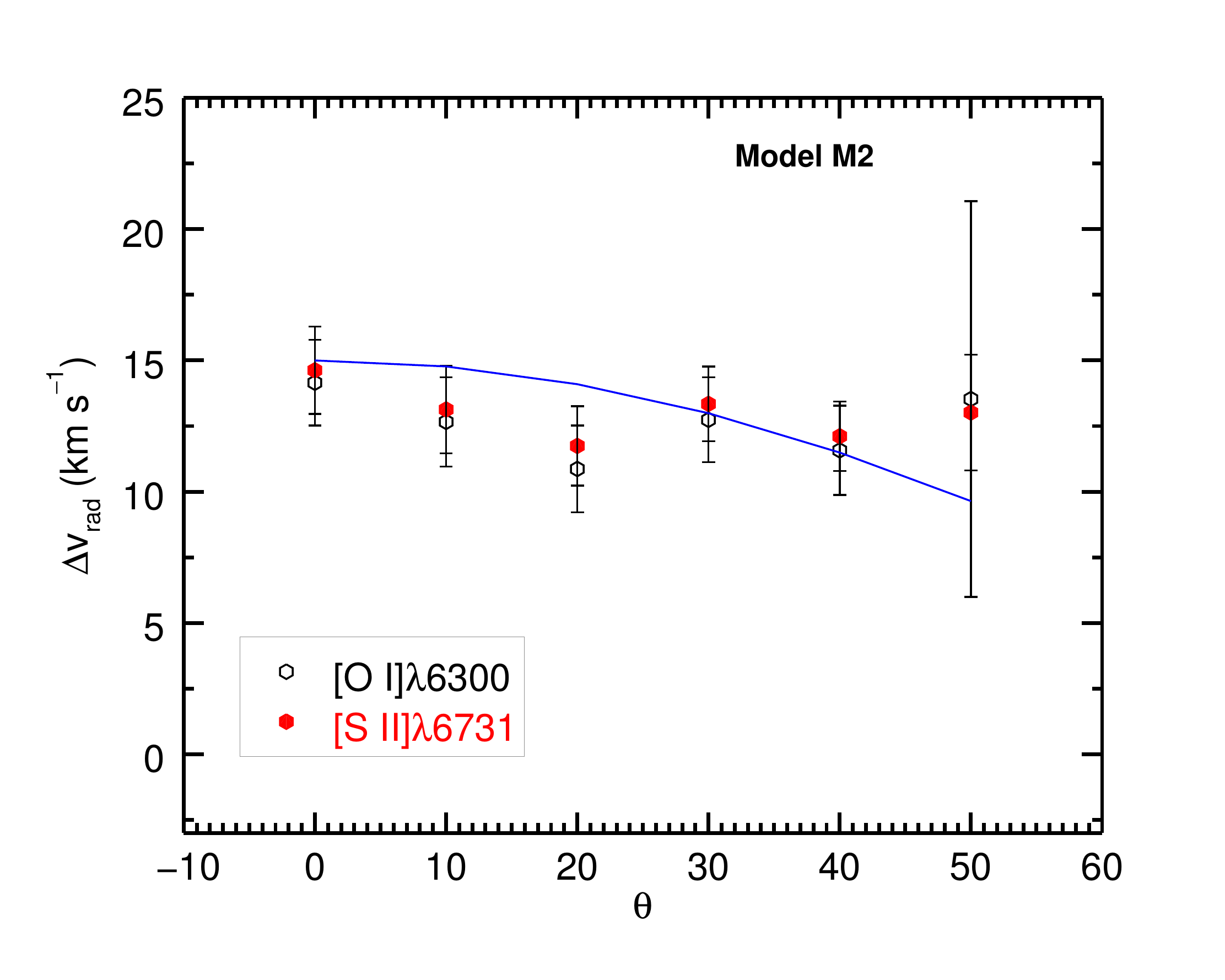}
\caption{The same as in Figure \ref{fig4}, but for M2 model (with a jet rotation velocity constant through the jet cross-section). Here,
the solid curve is a cosine function with amplitude of 15 km\,s$^{-1}$.
}
\label{fig5}
\end{figure}

In all models the VCMs were built first for the raw data and then
convolved with a profile in order to mimics the instrumental
effect on the data. For the convolution procedure we have assumed
a Gaussian profile, that goes to zero at 3 pixels of distance from
a given point in the map. At the distance of DG Tau, for instance, 
this corresponds to a point-spread function of $0.1$ arcsec
(of FWHM). 

The spectra extraction procedure has been already presented
in CER06 and is described in detail in the Appendix.
Briefly, we define four different regions along the jet
axis, and near the jet inlet. These regions have a superposition
of one cell along the jet propagation direction. For each one of
these regions, we define a $3\times 3$ square region for which a
mean spectra is taken (i.e., we sum up nine spectra to increase the signal). 
To build up the line profile, we use VCM from -400
km\,s$^{-1}$ to 100 km\,s$^{-1}$, with a 10 km\,s$^{-1}$ sampling.
We did the same procedure for different inclination angles, namely,
0, 10, 20, 30, 40 and 50 degrees.  The convolution procedure and
the differences in the inclination angles, as well as the presentation
of new models (M1 and M2) is the main difference of these results
when compared with those presented in CER06. 

In Figure \ref{fig4} we show the radial velocity shift for model
M1, taking into account the [\ion{S}{2}] and [\ion{O}{1}] emission
lines. 
The dispersion around the mean value is also plotted as vertical,
error bars.  The results for the non-rotating M1 model in Figure
\ref{fig4} is compared with the functions $20\, \sin \theta$ and
$10 \, \sin \theta$.  They are consistent with a sinusoidal fit,
with different amplitudes for the different emission lines.

\begin{figure}
\centering
\includegraphics[width=8cm]{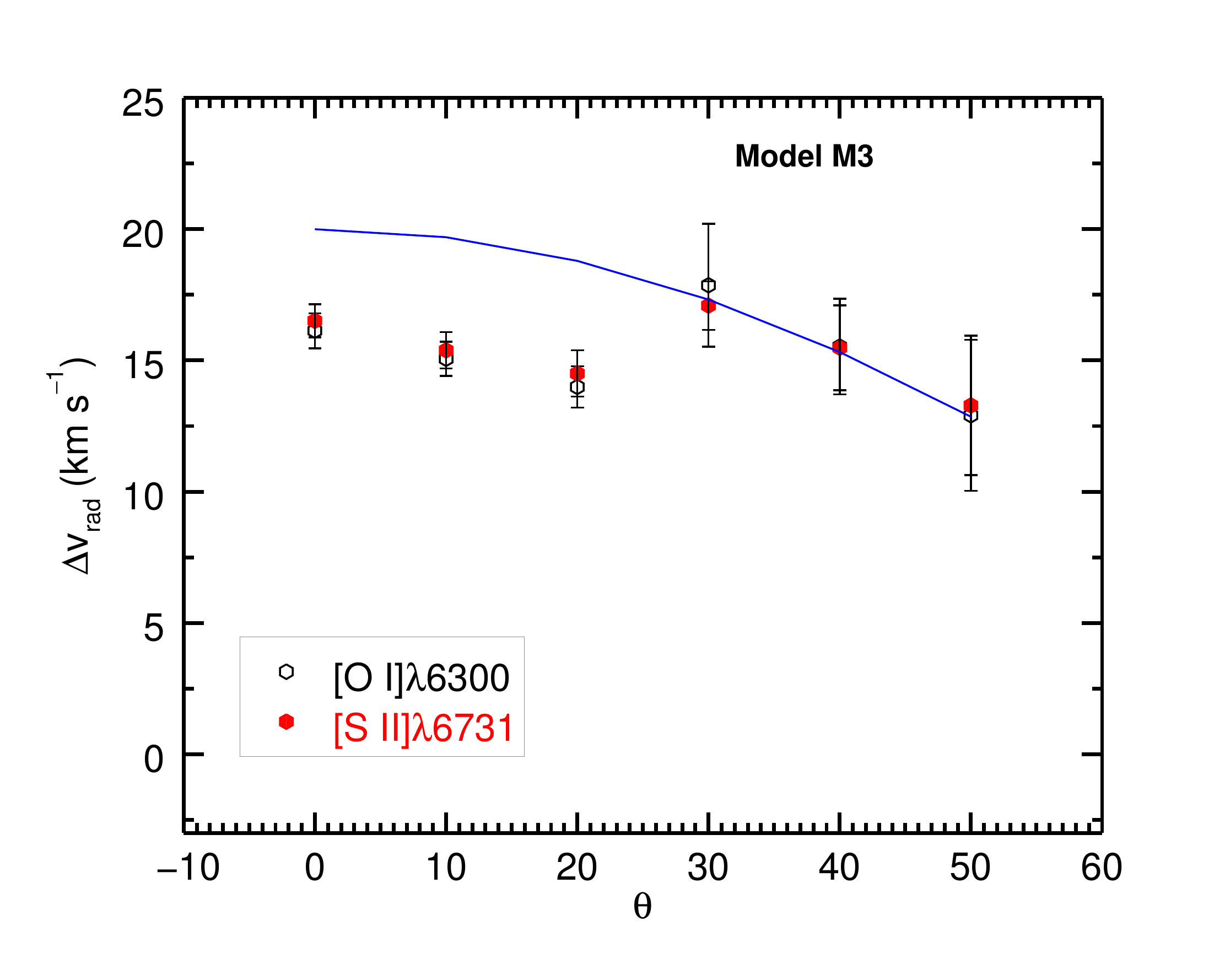}
\caption{The same as in Figure \ref{fig4}, but for M3 model (with a jet rotation velocity decreasing as $v_\phi \propto r^{-1}$). The
solid curve is a cosine function with amplitude of 20 km\,s$^{-1}$.}
\label{fig6}
\end{figure}

\begin{figure}
\centering
\includegraphics[width=8cm]{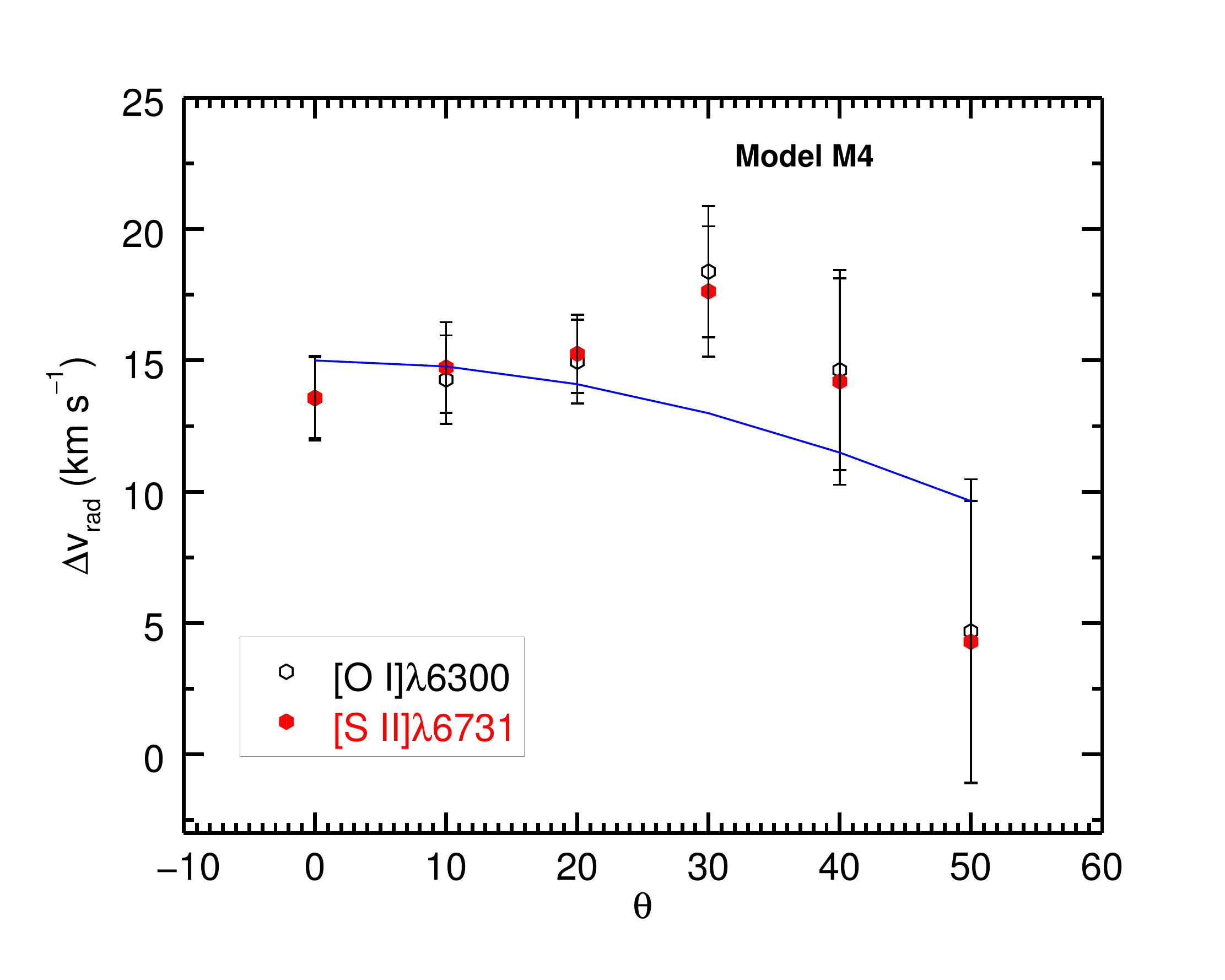}
\caption{The same as in Figure \ref{fig4}, but for M4 model (rotating,
precessing jet model). The solid curve is a cosine function with
amplitude of 15 km\,s$^{-1}$.}
\label{fig7}
\end{figure}

In Figure \ref{fig5} we show the same result for the (rotating) M2
model. Now the data are compared with the function $15 \cos \theta$. 
They are consistent with a cosine trend,
although slightly below it (for the chosen value of 15 km
s$^{-1}$ for the amplitude of the oscillation) except at large jet
inclination angles.  The same behavior
is also seen for the M3 model (see Figure \ref{fig6}).  A comparison
between figure \ref{fig5} and figure \ref{fig6} also shows that the
result is nearly independent (at least qualitatively) from the
particular dependence of the rotation velocity with radius considered.
In Figure \ref{fig7} we show the difference in the radial velocity
for the M4 (rotating and precessing model) as a function of the
inclination angle. Depicted also in this figure is the cosine
function (for the sake of comparison). 
Figures \ref{fig5}-\ref{fig7} all show an appreciable TVS.
These results are in agreement
with those presented by CER06, since their M2 and M4 models are the same
as the ones in the present paper. Although less evident, we should note that \cite{smith07} have also
found some TVS in their precession model of molecular jet (see their
Figure 14, Regions III and IV). Furthermore, the rotating models discussed
in \cite{smith07} and CER06 display signatures for rotation in the TVS analysis
that are not precisely equivalent. In this case, however, the
initial rotational profile seems to be at the origin of the differences reported,
as has been pointed out by \cite{smith07}, and the results obtained by
both are consistent with the expected ones taken into account
their adopted initial rotational profiles.

\section{Discussion}
\label{sec4}

In this paper, we analyze the origin of velocity shifts observed 
in protostellar jets. Three-dimensional numerical simulations 
presented in section \ref{sec3} confirm the results of section 
\ref{sec2}:  radial velocity shifts strongly depend on the jet 
inclination angle and scale with respect to rotation as 
$\sim \cos\theta$ and side to side velocity asymmetries as 
$\sim \sin\theta$.

An interesting difference between the results of the numerical
simulations and the observations is the behavior of the velocity
shifts for different emission lines. While in the simulations the
[\ion{O}{1}]$\lambda$6300 and the [\ion{S}{2}]$\lambda$6731 emission
lines have a similar behavior, in the observations by \citet{coffey04,
coffey07} the [\ion{O}{1}]$\lambda$6300 line exhibits velocity
shifts much larger than those observed in the [\ion{S}{2}]$\lambda$6731
emission line (but these two lines follow a similar behavior in
the radial shifts measured by \citealt{bacciotti02} for the DG\,Tau
micro-jet). 

Emission line intensities strongly depend on the jet density profile. 
The setup of our simulations considers the same abundance 
for a given atom (with respect to hydrogen) at the jet inlet. In addition,
in the simulations presented in this paper we have followed the 
``standard'' recipe of injecting a jet with a top-hat
density profile while a ``real'' jet has a density profile larger 
on the jet axis (see, e.g., the electron density profile of the HH\,30 jet
reconstructed by tomographic techniques by \citealt{decolle10}). 
Therefore, the adopted top-hat velocity and density
profiles can actually affect a direct comparison with
the observations.

One could think that the results presented in Section \ref{sec2} depend strongly 
on the HH\,30 data which, moving nearly in the plane of the sky and with very 
low values of  $\Delta v_r$, favor a model with low rotation and large velocity asymmetry.
This is actually not the case, as the results are not strongly dependent 
on the HH\,30 data. In fact, a fit computed without including 
the HH\,30 emission lines gives $v_\phi=5.1 \pm 0.7 $ km\,s$^{-1}$, 
$\Delta v_z = 12.2\pm 2.1$  km\,s$^{-1}$, and $Q=0.96$ for the 
[\ion{O}{1}] line, and $v_\phi=5.8\pm 2.2$ km\,s$^{-1}$, 
$\Delta v_z = 4.2\pm 6.3$  km\,s$^{-1}$, and $Q=0.40$ for the [\ion{N}{2}] 
line, then confirming that velocity asymmetries are an important component
of the radial velocity shifts, at least at large jet inclination angles.

As a consequence, jets with low inclination angles should be preferentially 
used to properly infer the amount of rotation occurring in the
velocity shifts. Among the jets considered here, HH\,30 is nearly 
in the plane of the sky but,
as mentioned above, does not show radial velocity shifts
larger than the statistical error. 
Th\,28 has a small inclination angle ($\theta \sim 10^\circ$) with respect
to the plane of the sky, and presents large velocity shifts.

As discussed in section \ref{sec2}, if a side to side spatial 
asymmetry is present in an emission line profile, it becomes very 
important to address the effect of the asymmetry (possibly due to 
an asymmetric shock) on the interpretation of the velocity shift as due to rotation.

In general large scale stellar jets are very asymmetric and irregular, 
while small scale jets exhibit some degree of symmetry.
Several jets among those used to measure rotation in jets 
show asymmetries in the physical parameters. \citet{coffey08}, 
using data from STIS, determined the physical parameters (electron and 
hydrogen density, temperature, and ionization fraction) from 
PV diagrams in the red-shifted RW\,Tau and Th\,28 jets, and in the 
blue-shifted DG\,Tau, HH\,30 and CW\,Tau jets. Among these jets, 
Th\,28, which, as mentioned above, 
is the most promising jet to infer rotation from radial velocity shift, 
presents a very clear side to side asymmetry in the electron density, 
while temperature and ionization fraction are nearly symmetric.  
The DG\,Tau jet presents also a strong asymmetry in the electron 
density, while HH\,30 and RW\,Aur are nearly symmetric (at least, at 
the position where the slit is located and with the resolution of STIS).
For the Th\,28 and DG\,Tau jets, the more obvious explanation for the 
origin of the asymmetry in the electron and total density is the
existence of an asymmetry in the jet velocity.

Several processes, which can create asymmetries in the jet,
have been extensively studied, mainly theoretically.
They can be of ``internal'' origin, i.e. due to a non-symmetric injection 
velocity from the star disk system \citep[e.g.,][]{soker05}
or of ``external'' origin, i.e. due to the jet environment, 
originated for instance by hydrodynamics of magnetohydrodynamics 
instabilities (e.g., ``kink'' modes), by lateral gradients in the density of 
a stratified interstellar medium \citep[e.g.,][]{canto96}, the presence of 
photoionization \citep[][]{bally06, masciadri01a}, a wind on one side of the jet 
\citep[e.g., ][]{canto95, masciadri01b, ciardi08},
or, in general, the motion of the source with respect to the environment.
For example, the HH\,30 jet shows a small bending from distances of
order of 400 AU from the central source \citep{anglada07}.
As the effect needed to explain the observed velocity shift is quite small
($\sim 10$\% of the observed poloidal jet velocity), the required amount
of asymmetry is often smaller than the one discussed in the cited 
papers.

Several authors \citep[e.g.,][]{anderson03, ferreira06, pesenti04} have 
considered, from the observation of transverse shifts, the implications 
on the jet ejection models, showing that the observation are consistent 
with the disk-wind ejection mode. 
Our results imply that  jet rotation, if present, is 
probably smaller than the values inferred by previous authors. 
That does not  imply  necessarily that the rotation at the base of the 
jet is small. Shocks, jet expansion, entrainment, can all potentially
lead to a ``loss of memory'' of the material with respect to the original 
rotation, as discussed for instance by \citet{fendt11}
for magnetohydrodynamics shocks in a helical magnetic field.
Recent numerical simulations \citep{staff15} showed 
that the signature of rotation in the jet can be even non-Keplerian, 
which means that the signature showed by RW Aur, in which the jet 
seems to rotate in the sense contrary to the disk rotation can, in fact, 
occur. This pose an additional difficulty, in our view, to interpret 
side to side differences in radial velocity as rotation, since we need 
to trace back the phase of the torsional Alfv\`en wave that is actually 
producing the outflow at a given point in the jet. 

Observations of molecular jets in some cases also show small velocity shifts ($\lesssim$ a few km s$^{-1}$) which can be interpreted as jet rotation. Although in this paper we have limited our analysis to atomic jets, our conclusions are also applicable to molecular jets, where the rotation features are observed at larger distances from the disk-star system  (both along and across the jet axis), i.e. at the ``edge'' of the jet, where the interaction with the environment is expected to be more important. \citet{zapata15} have recently suggested that observed TVS could be originated by rotation of the entrained material.


\section{Conclusion}
\label{sec5}

In this paper, we have discussed the uncertainties present 
when interpreting as rotation the velocity shifts observed in atomic
protostellar jets. We have shown that asymmetric shocks, possibly produced 
as the results of the interaction with the environment or by asymmetries 
in the ejection velocity from the disk-star system, may produce effects 
similar to those produced by rotation. We have also quantified, 
analytically and by detailed numerical simulations, how in jets with 
large inclination angles transverse velocity shifts may be dominated 
by jet asymmetries, while rotation, if present, should dominate in jets with 
low inclination angles. 

The main uncertainties in this study reside in the low statistics existing
in our analysis (only six jets) and on the fact that we are assuming that 
rotation and velocity asymmetry are the same for all jets.
The analysis presented here does not pretend to be complete, as
other factors (e.g., the agreement between the sense of rotation of jet and disk, 
the poloidal extension of the region showing velocity shifts) should be considered when analyzing the observed
TVS. Nevertheless, by analyzing existing data of a limited sample 
of atomic protostellar jets, we have clearly shown that velocity asymmetries
(whatever is their origin) seem to play a very important role in determining 
the amount of transverse velocity shift present in protostellar jets.

Our results imply that only a detailed modeling of the observations,
which include all the effects that can potentially play a role in generating
velocity shift, combined possibly with numerical simulations or with novel 
analysis of the data \citep[e.g., the 
principal component analysis suggested by][]{cerqueira15}, should
be used to properly determine the presence of rotation. Additionally, 
the study of rotation has to proceed carefully in jets that 
present clear side-to-side asymmetries in line emissivity profiles 
or in the physical parameters. Jets with low inclination angles and 
without large asymmetries (e.g., HH\,30) are the ideal candidate 
to be used to determine an upper limit on 
the rotation, the angular momentum transfer, and, from there, 
help distinguishing among different jet ejection models.

Our results do now imply that there is not rotation in protostellar jets. 
On one side, the rotation can be smaller than expected and/or 
dominated by shock asymmetries. On the other side, magnetic shocks, 
jet expansion, entrainment, among other phenomena, all lead to 
a ``loss of memory'' of the initial rotation of the jet material as it 
expands to large distances from  the jet ejection region.


\acknowledgments
We thank J. Cant\'o, S. Lizano and A.C. Raga for useful discussions. 
AHC thanks Cnpq/CAPES for financial support using the PROCAD
project (552236/2011-0) and CAPES/CNPq Science without Borders
program (under grants 2168/13-8). AR acknowledges Spanish MICINN grants 
AYA2011-30228-C03 and AYA2014-57369-C3-2-P (cofunded with FEDER funds),
and FDC the UNAM-PAPIIT grants IA103315, IG100516 and the HST Cycle 19 archival 
proposal 12633.


\appendix

\section{The fitting process}

\begin{figure}
\centering
\includegraphics[width=8.7cm]{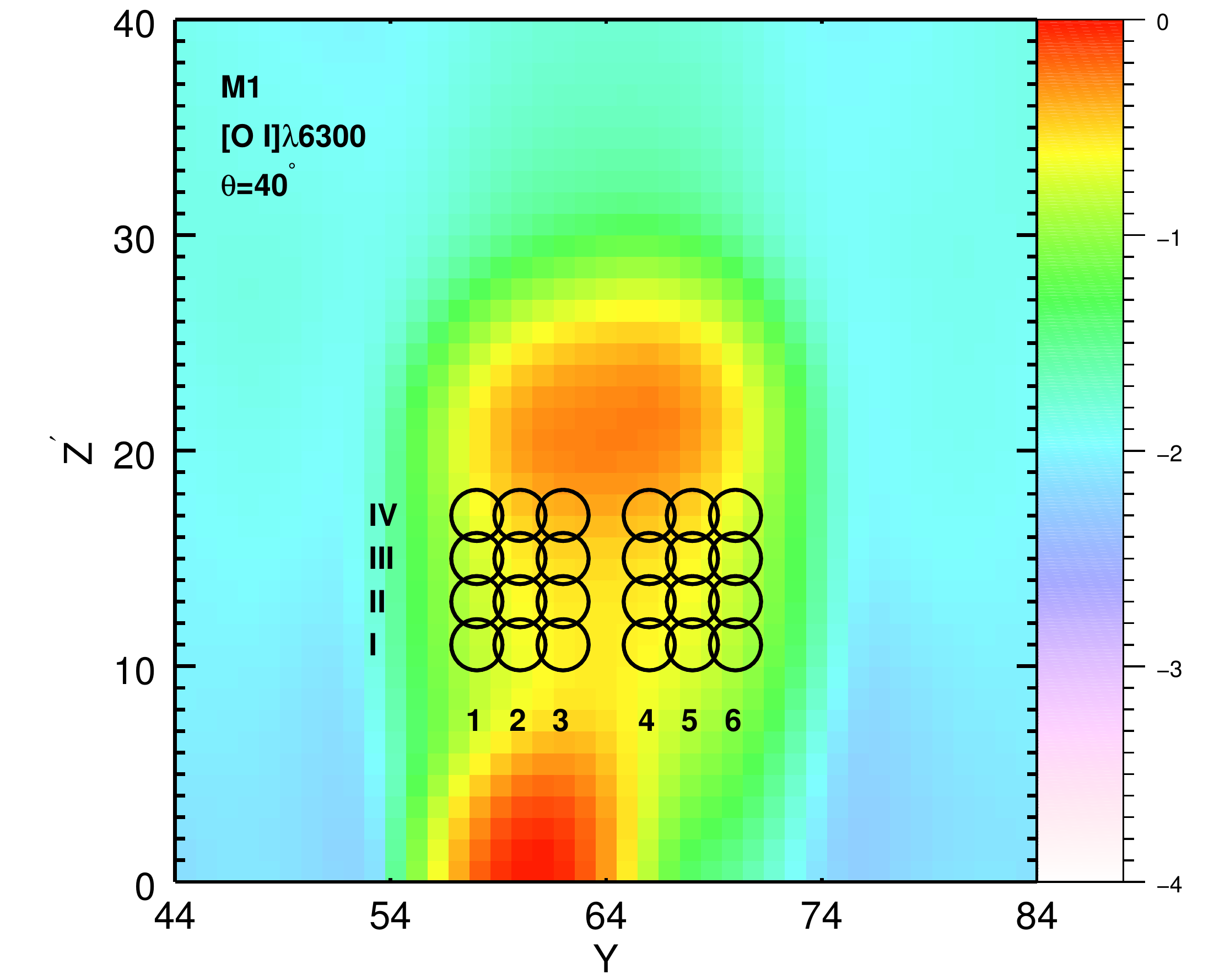}
\includegraphics[width=8.7cm]{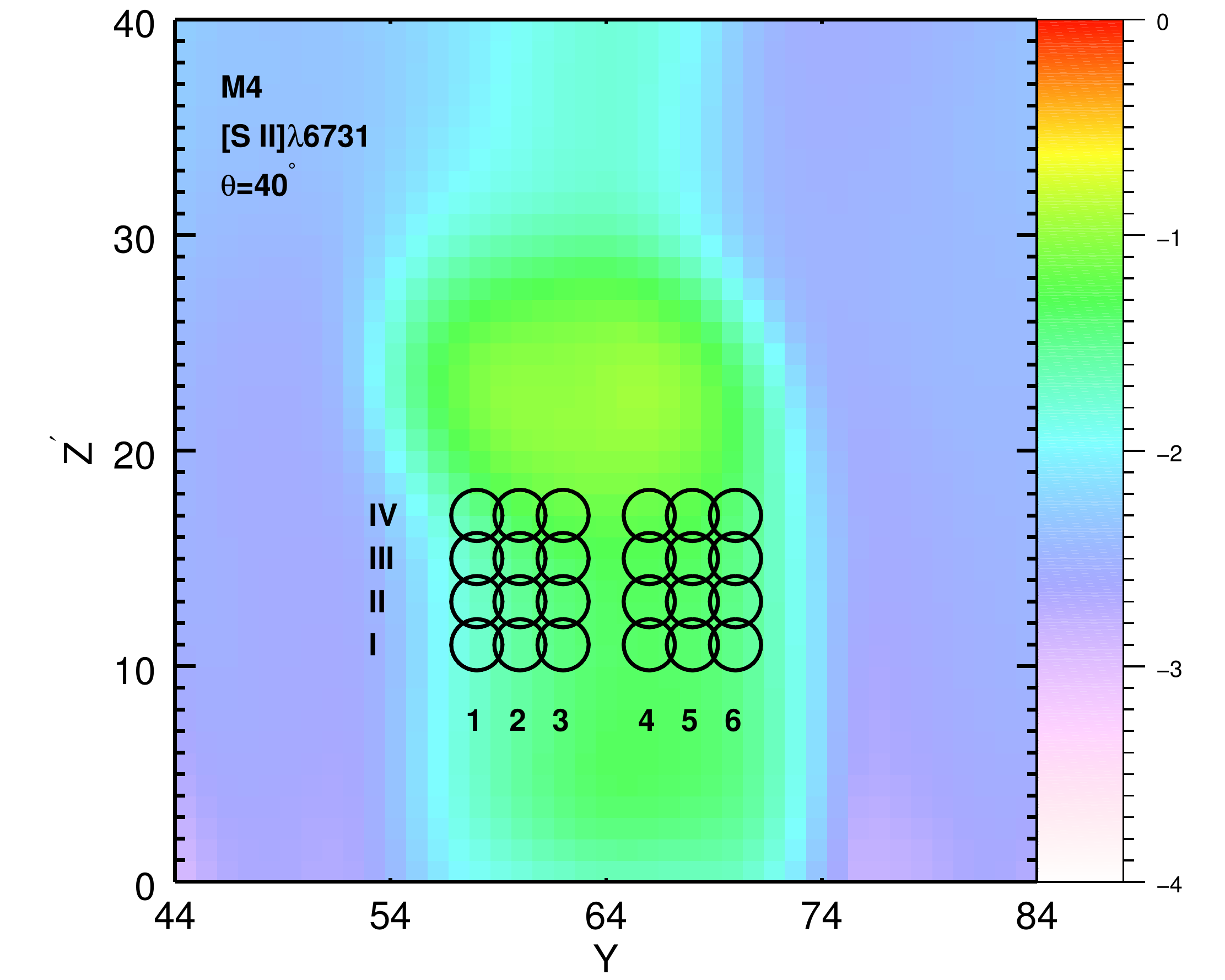}
\caption{Normalized and integrated emission line maps for model M1 (left) and M4 (right)
for two different lines, [\ion{O}{1}]$\lambda6300$ (left) and [\ion{S}{2}]$\lambda6731$ (right). The
jet is inclined $\theta = 40^{\circ}$ with respect to the line of sight
and the $Y$ and $Z^{\prime}$ coordinates are in code units. The jet inlet
is centered at $(y,z) = (64,0)$. In both cases we have superimposed the
slit positions (from 1 to 6), symmetrically disposed in both sides of
the jet axis, and the four regions (from I to IV).}
\label{figA1}
\end{figure}

Figure \ref{figA1} shows the integrated emission maps for
models M1 at [\ion{O}{1}]$\lambda6300$ (left) and for  M4 (right) at [\ion{S}{2}]$\lambda6731$, 
for a jet inclination angle (with respect to the line of sight) $\theta = 40^{\circ}$. The jet inlet is at $(y,z) =
(64,0)$\footnote{In code units of distance, or cell number. We note
that, since the maps in Figure \ref{figA1} represent the system
already inclined toward the observer, the Z$^{\prime}$ coordinate
represents, then, the projected distance in the ``plane of the sky''.}.
The slits are placed near the jet origin (as in CER06), but in
region IV the slits are already near the first internal working
surface (see the [\ion{O}{1}] map on the left side of Figure \ref{figA1}).

For a given emission line and inclination angle, we have
calculated the radial velocity shift considering six positions, symmetrically
disposed with respect to the jet axis, and in four different regions distributed
along the jet axis  (circles in both panels
in Figure \ref{figA1} indicate the precise position of these slits
and give an approximated idea of their aperture size, which is
actually $3\times3$ pixels). For each one of these slits, two
gaussians are adjusted to the line profile (a low/moderate velocity
component and a high velocity component), and the velocity differences
between both sides of the jet axis are calculated using the same gaussian
component (in particular, we have used the high-velocity component). 
A mean value is then calculated considering the different
regions (I, II, III and IV) as well as the different slit pairs (S1-S6,
S2-S5, S3-S4) for a given model, inclination angle
and emission line.

\begin{figure}
\centering
\includegraphics[width=8cm]{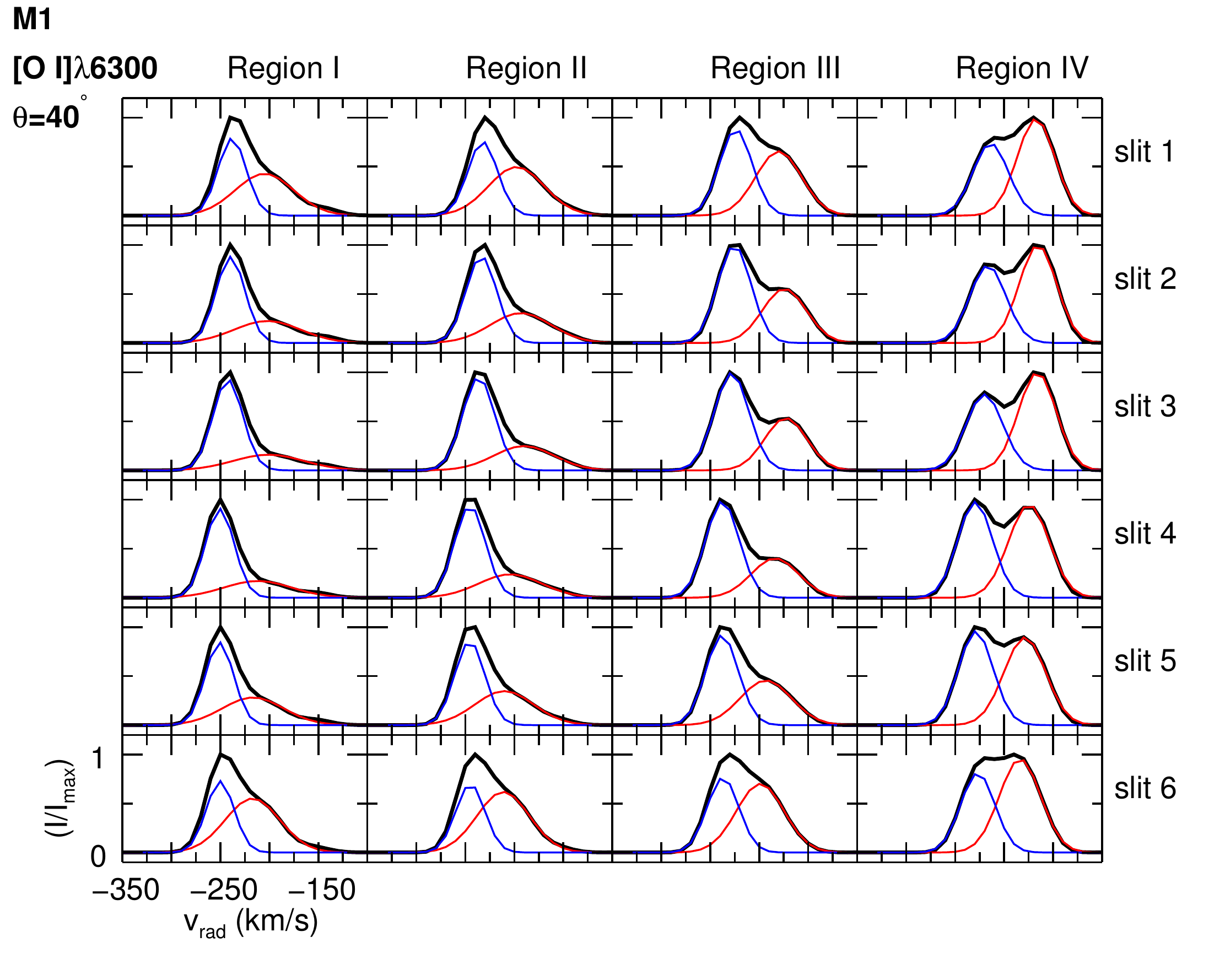}
\includegraphics[width=8cm]{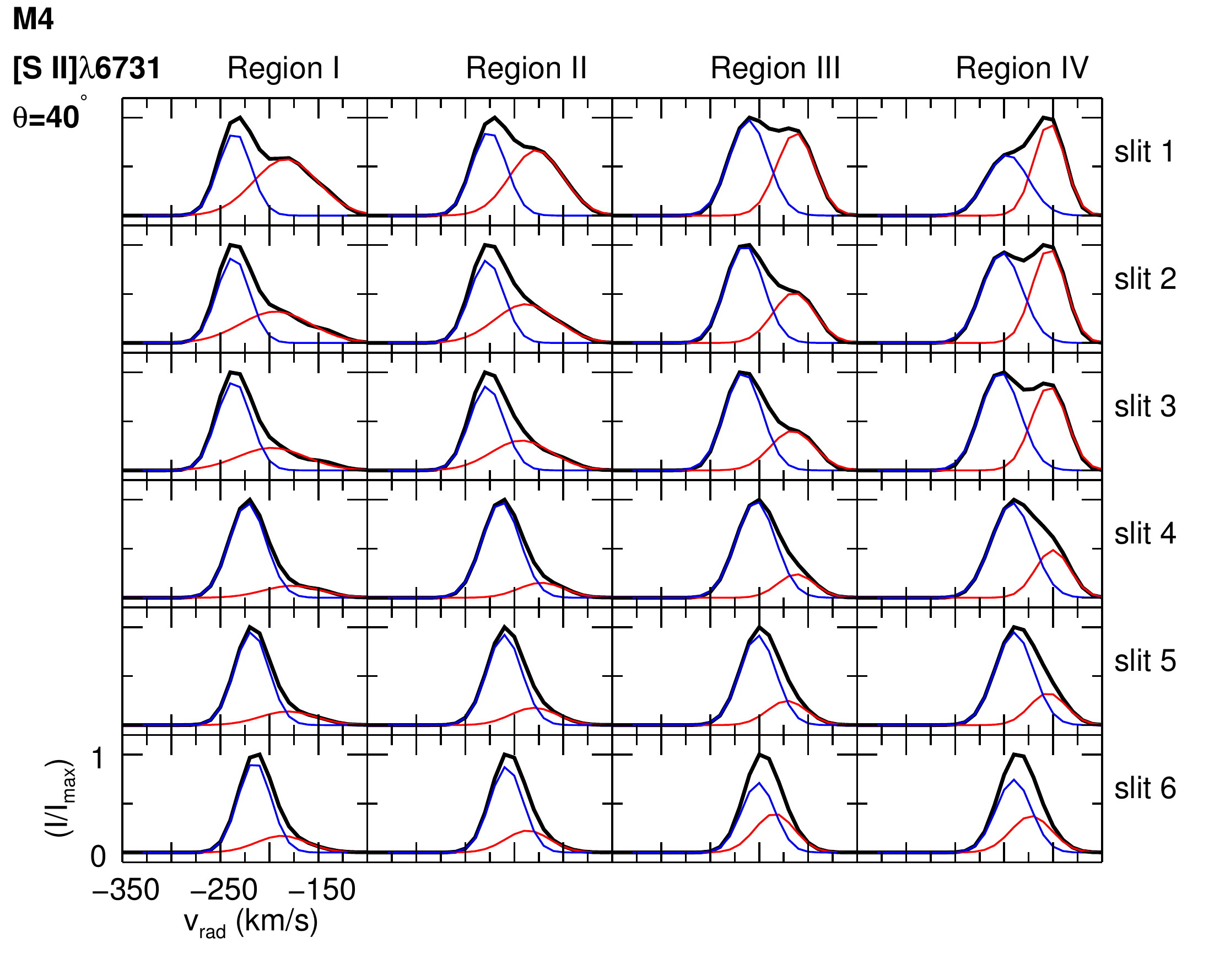}
\caption{Line profiles (black solid line) extracted for each
 slit (slits 1 to 6  from top to bottom, as indicated in the
rightmost part of the panels) defined in Figure \ref{figA1}, as
well as for each one of the four different regions (from I to IV;
from left to right as indicated in the topmost part of the panels)
for model M1 (top) and M4 (bottom), for the [\ion{O}{1}]$\lambda$6300 (top)
and [\ion{S}{2}]$\lambda6731$ (bottom) line profiles (see also Figure \ref{figA2}). Red
and blue solid lines represent the profiles for the adjusted
gaussians. The intensity in each spectra is normalized to its maximum
value and the velocity ranges from -350 km s$^{-1}$ to -100 km$^{-1}$
(see the left-bottom panel).}
\label{figA2}
\end{figure}

The extracted spectrum (black solid
lines) can be seen in Figure \ref{figA2}. The adjusted gaussians
have also been plotted in Figure \ref{figA2} (solid red and blue lines).
It is clearly seen that the profiles change from
slightly to highly asymmetrical, from the jet inlet (region I) towards
the internal working surface (region IV). We can also see that the
rotation changes the amplitude and velocity of the peak of the 
high-velocity component (see the shift in the peak positions in 
the profiles of Figure \ref{figA2}).

\begin{thebibliography}{}

\bibitem[Anglada et al.(2007)]{anglada07} Anglada, G., L{\'o}pez, R., Estalella, R., et al.\ 2007, \aj, 133, 2799 
\bibitem[Anderson et al.(2003)]{anderson03} Anderson, J.M. , Li, Z.-Y., Krasnopolsky, R., \& Blandford, R.D., 2003, ApJ, 590, 107
\bibitem[Bacciotti et al.(2002)]{bacciotti02} Bacciotti, F. , Ray, T.P., Mundt, R., Eisloffel, J., \& Solf, J., 2002, ApJ, 576, 222
\bibitem[Bally et al.(2006)]{bally06} Bally, J., Licht, D., Smith, N., \& Walawender, J.\ 2006, \aj, 131, 473 
\bibitem[Biretta et al.(2016)]{biretta16} Biretta, J. et al. 2016, in STIS Instrument Handbook, Version 15.0, (Baltimore: STScI).
\bibitem[Blandford \& Payne(1982)]{blandford82} Blandford, R.~D., \& Payne, D.~G.\ 1982, \mnras, 199, 883 
\bibitem[Cabrit et al.(2006)]{cabrit06} Cabrit, S., Pety, J., Pesenti, N., \& Dougados, C., 2006, A\&A,452, 897
\bibitem[Canto \& Raga(1995)]{canto95} Canto, J., \& Raga, A.~C.\ 1995, \mnras, 277, 1120 
\bibitem[Canto \& Raga(1996)]{canto96} Canto, J., \& Raga, A.~C.\ 1996, \mnras, 280, 559 
\bibitem[Cerqueira \& de Gouveia Dal Pino(2004)]{cerqueira04} Cerqueira, A.H., \& de Gouveia dal Pino, E.M. 2004, A\&A, 426, L25
\bibitem[Cerqueira et al.(2006)]{cerqueira06} Cerqueira, A.H., Vel\'azquez, P.F., Raga, A.C., Vasconcelos, M.J., \& De Colle, F., 2006, A\&A, 448, 231
\bibitem[Cerqueira et al.(2015)]{cerqueira15} Cerqueira, A.~H., Reyes-Iturbide, J., De Colle, F., \& Vasconcelos, M.~J.\ 2015, \aj, 150, 45
\bibitem[Chrysostomou et al.(2008)]{chrysostomou08} Chrysostomou, A., Bacciotti, F., Nisini, B., Ray, T.P., Eisloffel, J., Davis, C.J., \& Takami, M., 2008, A\&A 482, 575
\bibitem[Choi et al.(2011)]{choi11} Choi M., Kang, M., \& Tatematsu, K. 2011, \apj, 728, L34
\bibitem[Ciardi et al.(2008)]{ciardi08} Ciardi, A., Ampleford, D.~J., Lebedev, S.~V., \& Stehle, C.\ 2008, \apj, 678, 968-973 
\bibitem[Codella et al.(2004)]{codella07} Codella, C., Cabrit, S., Gueth, F., Cesaroni, R., Bacciotti, F., Lefloch, B., \& McCaughrean, M.J., 2007, A\&A, 462, L53
\bibitem[Coffey et al.(2004)]{coffey04} Coffey, D., Bacciotti, F., Woitas, J., Ray, T.P., \& Eisloffel, J., 2004, ApJ, 604, 758
\bibitem[Coffey et al.(2007)]{coffey07} Coffey, D., Bacciotti, F., Ray, T.P., Eisloffel, J., \& Woitas, J., 2007, ApJ, 663, 350
\bibitem[Coffey et al.(2008)]{coffey08} Coffey, D., Bacciotti, F., \& Podio, L., 2008, ApJ, 689, 1112
\bibitem[Coffey et al.(2011)]{coffey11} Coffey, D., Bacciotti, F., Chrysostomou, A., Nisini, B., \& Davis, C. 2011, \aap, 526, 40
\bibitem[Coffey et al.(2012)]{coffey12} Coffey, D., Rigliaco, E., Bacciotti, F., Ray, T.~P., \& Eisl{\"o}ffel, J.\ 2012, \apj, 749, 139 
\bibitem[Coffey et al.(2015)]{coffey15} Coffey, D., Dougados, C., Cabrit, S., Pety, J., \& Bacciotti, F.\ 2015, \apj, 804, 2 
\bibitem[Davis et al.(2000)]{davis00} Davis, C.J., Berndsen, A., Smith, M.D., Chrysostomou, A., \& Hobson, J., 2000, MNRAS 314, 241
\bibitem[De Colle et al.(2008)]{decolle08} De Colle, F., del Burgo, C., Raga, A.C., 2008, A\&A, 485, 765
\bibitem[De Colle et al.(2010)]{decolle10} De Colle, F., del Burgo, C., \& Raga, A.~C.\ 2010, \apj, 721, 929  
\bibitem[Dougados et al.(2000)]{dougados00} Dougados, C., Cabrit, S., Lavalley, C., M\'enard, F. 2000, \aap, 357, L61
\bibitem[Fendt(2011)]{fendt11} Fendt, C.\ 2011, \apj, 737, 43 
\bibitem[Ferreira et al.(2006)]{ferreira06} Ferreira, J., Dougados, C., \& Cabrit, S., 2006, A\&A 453, 785–796
\bibitem[Hartigan \& Hillenbrand(2009)]{hartigan09} Hartigan, P., \& Hillenbrand, L. 2009, \apj, 705, 1388
\bibitem[Lavalley-Fouquet et al.(2000)]{lavalley00} Lavalley-Fouquet, C., Cabrit, S., \& Dougados, C.\ 2000, \aap, 356, L41 
\bibitem[Launhardt et al.(2009)]{laun09}  Launhardt, R., Pavlyuchenkov, Ya., Gueth, F., Chen, X., Dutrey, A., Guilloteau, S., Henning, Th., Pi\'etu, V., Schreyer, K., Semenov, D. 2009, A\&A, 494, 147
\bibitem[Lee et al.(2007)]{lee07} Lee, C.-F., Ho, P.T.P., Palau, A., Hirano, N., Bourke, T.L., Shang, H., \& Zhang, Q., 2007, ApJ, 670, 1188
\bibitem[Lee et al.(2008)]{lee08} Lee, C.-F., Ho, P.T.P., Bourke, T.L., Hirano, N., Shang, H., \& Zhang, Q., 2008, ApJ, 685, 1026
\bibitem[Li et al.(2014)]{li14} Li, Z-Y., Banerjee, R., Pudritz, R.E. et al. 2014, in Protostars and Planets VI, University of Arizona Press (2014), eds. H. Beuther, R. Klessen, C. Dullemond, Th. Henning
\bibitem[Livio(1999)]{livio99} Livio, M. 1999, Physics Reports, 331 (3-5), 225
\bibitem[Louvet et al.(2016)]{louvet16} Louvet, F., Dougados, C., Cabrit, S., et al.\ 2016, arXiv:1607.08645 
\bibitem[Masciadri \& Raga(2001a)]{masciadri01a} Masciadri, E., \& Raga, A.~C.\ 2001, \aap, 376, 1073 
\bibitem[Masciadri \& Raga(2001b)]{masciadri01b} Masciadri, E., \& Raga, A.~C.\ 2001, \aj, 121, 408 
\bibitem[Pech et al.(2012)]{pech2012} Pech, G., Zapata, L.A., Loinard, L., \& Rodr\'{\i}guez, L.F. 2012, \apj, 751, 78
\bibitem[Pesenti et al.(2004)]{pesenti04} Pesenti, N., Dougados, C., Cabrit, S., Ferreira, J., Casse, F., Garcia, P., \& O’Brien, D., 2004, A\&A, 416, L9
\bibitem[Pety et al.(2006)]{pety06} Pety, J., Gueth, F., Guilloteau, S., \& Dutrey, A., 2006, A\&A, 458, 841
\bibitem[Pudritz et al.(2007)]{pudritz07} Pudritz, R.E., Ouyed, R., Fendt, C., \& Brandenburg, A., 2007, Protostars and Planets V, 277
\bibitem[Raga, Navarro-Gonz\'alez \& Villagr\'an-Muniz(2000)]{raga00} Raga, A.C., Navarro-Gonz\'alez, R. \& Villagr\'an-Muniz, M. 2000, RMxAA, 36, 67
\bibitem[Raga et al.(2001)]{raga01} Raga, A., Cabrit, S., Dougados, C., \& Lavalley, C.\ 2001, \aap, 367, 959 
\bibitem[Raga et al.(2004)]{raga04} Raga, A.C., Riera, A., Masciadri, E., Beck, T., B\"ohm, K.H., \& Binette, L. 2004, AJ, 127, 1081
\bibitem[Raga et al.(2007)]{raga07} Raga, A.C., De Colle, F., Kajdic, P., Esquivel, A., \& Cant\'o, J. 2007, A\&A, 465, 879
\bibitem[Reipurth \& Bally(2001)]{reipurth01} Reipurth, B., \& Bally, J.\ 2001, \araa, 39, 403 
\bibitem[Shu et al.(2000)]{shu00} Shu, F.H., Najita, J., Shang, H., \& Li, Z.-Y., 2000, Protostars and Planets IV, University of Arizona Press, Eds. V. Mannings, A.P. Boss, 789
\bibitem[Soker(2005)]{soker05} Soker, N., 2005, A\&A, 435, 125
\bibitem[Smith \& Rosen(2007)]{smith07} Smith, M.D., Rosen, A., 2007, MNRAS, 378, 691
\bibitem[Staff et al.(2015)]{staff15} Staff, J.E., Koning, N., Ouyed, R. et al. 2015, MNRAS, 446, 3975
\bibitem[van Albada, van Leer \& Roberts(1982)]{vanalbada82} van Albada, G.D., van Leer, B., \& Roberts, W.W.Jr. 1982, A\&A, 108, 76
\bibitem[Yuan \& Narayan(2014)]{yuan14} Yuan, F. \&  Narayan, R. 2014, ARA\&A, 52, 529
\bibitem[Woitas et al(2005)]{woitas05} Woitas, J., Bacciotti, F., Ray, T.P., Marconi, A., Coffey, D., \& Eisloffel, J., 2005, A\&A, 432, 149
\bibitem[Zapata et al.(2015)]{zapata15} Zapata, L.~A., Lizano, S., Rodr{\'{\i}}guez, L.~F.,  et al.\ 2015, \apj, 798, 131

\end{thebibliography}
\end{document}